\documentclass{iau}
\usepackage{graphicx}

\title[Observations of asteroid 4179 Toutatis] 
{Chang'e-2 spacecraft observations of asteroid 4179 Toutatis}

\author[Jianghui Ji et al.]   
{Jianghui Ji$^1$, Yun Jiang$^1$, Yuhui Zhao$^1$, Su Wang$^1$, Liangliang Yu$^{1,2}$
}

\affiliation{$^1$Key Laboratory of Planetary Sciences, Purple
Mountain
Observatory, \\
Chinese Academy of Sciences, Nanjing 210008, China  \\
$^2$Lunar and Planetary Science Laboratory, Macau University of Science and Technology, Taipa, Macau \\
email: {\tt jijh@pmo.ac.cn} }

\pubyear{2015}
\volume{318}  
\jname{Asteroids: New Observations, New Models}
\editors{Steven Chesley, Alessandro Morbidelli \& Robert Jedicke, eds.}

\begin{document}

\maketitle

\begin{abstract}
On 13 December 2012, Chang'e-2 completed a
successful flyby of the near-Earth asteroid 4179 Toutatis at a
closest distance of 770 meters from the asteroid's surface. The
observations show that Toutatis has an irregular surface and its
shape resembles a ginger-root of a smaller lobe (head) and a larger
lobe (body). Such bilobate shape is indicative of a contact binary
origin for Toutatis. In addition, the high-resolution images better
than 3 meters provide a number of new discoveries about this
asteroid, such as an 800-meter depression at the end of the large
lobe, a sharply perpendicular silhouette near the neck region,
boulders, indicating that Toutatis is probably a rubble-pile
asteroid. Chang'e-2 observations have significantly revealed new
insights into the geological features and the formation and
evolution of this asteroid. In final, we brief the future Chinese
asteroid mission concept.

\end{abstract}

\firstsection 
\section{Introduction}

Asteroid 4179 Toutatis was first discovered in 1934 and observed
once again in 1989. The asteroid is an Apollo-type Near-Earth Object
(NEO) that moves on a near 1:4 resonant orbit with Earth. Thus,
ground-based telescopes, especially radar facilities, performed
extensive observations when Toutatis approached Earth every four
years since 1992. Radar observations acquired from Arecibo and
Goldstone during the past two decades show that Toutatis bears an
irregular shape with two distinct lobes. Optical and radar
measurements further reveal that Toutatis is a non-principal axis
(NPA) rotating asteroid, which may result from Earth-approaching
flybys in the dynamical evolution
(\cite{Ostro95,Hudson98,Ostro99,Ostro02,Hudson03,Busch11}).

Using higher resolution delay-Doppler radar observations,
\cite[Hudson \& Ostro (1995)]{Hudson95}, Hudson et al. (2003) and
\cite[Busch et al. (2012)]{Busch12} established Toutatis'
3-dimensional shape models. The dimensions along three principal
axes were determined to be 1.92, 2.40 and 4.60 kilometers,
respectively. Toutatis was believed to rotate around long-axis with
a period of 5.41 days and the long-axis precesses with a period of
7.35 days (\cite{Busch10}). Moment of inertia ratios  were evaluated
to be $3.22\pm0.01$ and $3.09\pm0.01$ (\cite{Ostro99}). Recently,
\cite[Takahashi et al. (2013)]{Takahashi13} modeled the rotational
dynamics and evaluated the spin state parameters of Toutatis with
radar data spanning from 1992 to 2008, and they showed that the
solar and terrestrial gravitational tidal torques can play a role in
affecting its angular momentum.

\begin{figure*}
\begin{center}
\includegraphics[scale=0.43]{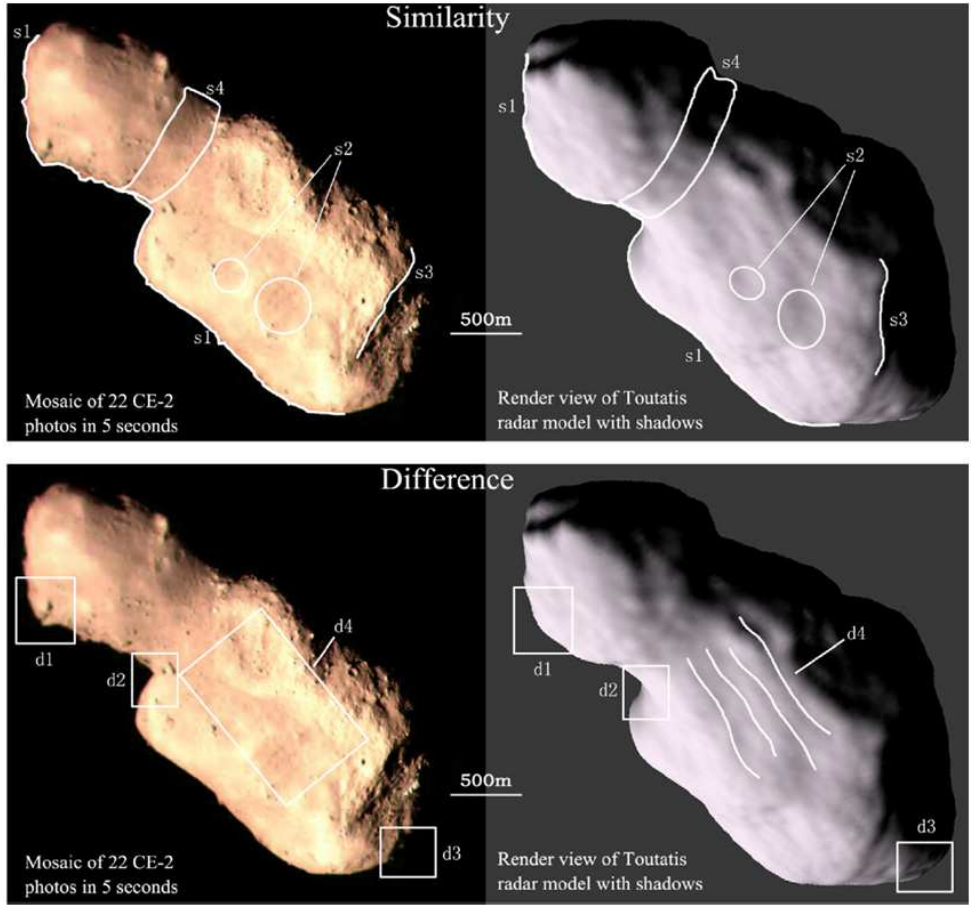}
\caption{Comparison of Toutatis' geological features between optical
images acquired by Chang'e-2 (color, left) and the radar model
(gray, right)  by Hudson et al. (2003). Similarities (s1-s4) and
differences (d1-d4) are marked. This figure is reproduced from Zou
et al. (2014).} \label{geo1}
\end{center}
\end{figure*}

As an NEO originates from the main belt, Toutatis, which was a
suitable target for the Chang'e-2 flyby mission, may provide key
clues on the formation of the Solar system. In addition, Toutatis
appears to be the largest Potential Hazardous Asteroids (PHAs) which
is full of enigmas - it is a NPA rotator with a bilobate shape
(\cite{Ostro95, Hudson03}) and may have a rubble pile structure. The
flyby mission confirmed the radar model's double-lobed appearance of
Toutatis, and optical images acquired provide new insights on the
origin and evolution of the asteroid (\cite{Huang13}).

Chang'e-2, as the second Chinese spacecraft to the Moon's
exploration, was launched on 1 October 2010. During the mission, the
probe orbited the Moon for six months. High-resolution images of the
lunar surface were obtained for studying Moon's morphology. After
the successful mission around the Moon, Chang'e-2 departed its orbit
on 9 June 2011 and moved its way to the Sun-Earth Lagrangian point
(L2) for exploring the space environment. It arrived L2 on 25 August
2011. Subsequently, on 12 December 2012, the asteroid would move to
the closest approach to the Earth. After over 230 days stay at L2,
Chang'e-2 started its mission to Toutatis\footnote{Before the flyby
mission, the ground-based observation campaign for Toutatis was
sponsored by the Chinese Academy of Sciences from May to November in
2012. We collected the observations from the Minor Planet Center as
well as those from the observation campaign to refine the orbit of
Toutatis, with uncertainties on the order of a few kilometers
(\cite{Huang13, Huang2013}).} on 1 June 2012 and on 13 December 2012
the spacecraft had a closest miss at about 770 $\pm$ 120 (3$\sigma$)
meters from Toutatis' surface (Huang et al. 2013a). It was the first
time that the images of Toutatis were acquired so closely. There
were about 425 images obtained. The highest resolution images of
Toutatis are better than 3 $\rm{m~ pixel^{-1}}$. Now Chang'e-2 is
still alive and flying to the space far away from Earth, more than
100 million kilometers in a heliocentric orbit.

In this work, we review the results of Toutatis as observed from the
Chang'e-2 flyby mission. Section 2 describes the surface geological
features of Toutatis. The orientation and rotational parameters of
Toutatis are determined in combination of the optical images of
Chang'e-2 and radar measurements in Section 3. The formation
scenarios are discussed in Section 4.  In final, future Chinese
asteroid mission is introduced in Section 5.

\begin{figure*}
\begin{center}
\includegraphics[scale=0.41]{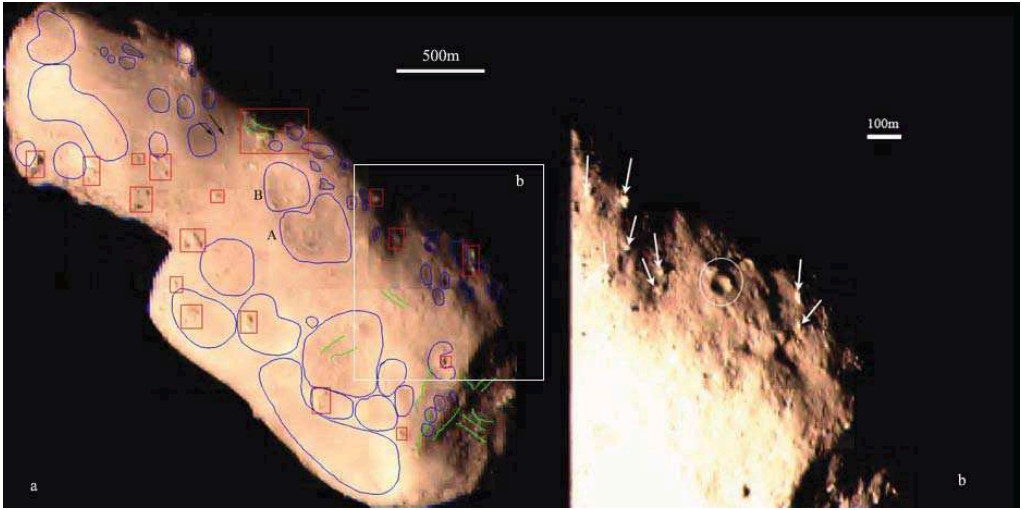}
\caption{Various geological features on the surface of Toutatis. (a)
Craters (blue circles), boulders (red squares), lineaments (green
lines) as well as the flow direction of regolith (black arrows) are
outlined. (b) A morphological-integrity crater shows a sharp bowl
shape, with dozens of boulders distributing around. This figure is
reproduced from Huang et al. (2013a).} \label{geo2}
\end{center}
\end{figure*}

\section{Surface Geological Features}

The Chang'e-2 flyby allows for observing the morphology of $\sim$
45\% surface of Toutatis. Figure \ref{geo1} shows that both the
optical images acquired by Chang'e-2 and the radar-derived shape
model exhibit similar silhouette (s1), two circular concavities in
the middle area of body (s2), a scarp (possibly crater rim) with
similar length and slope (s3) and the joint part with similar width,
shape and location (s4)(Zou et al. 2014). However, several
differences also exist: d1. A small angular feature appears visible
on the profile of head in the flyby image, which is vaguely shown in
the radar model. d2. A perpendicular profile near the neck region is
sharp in the image but vague in the model. d3. The far side of the
big lobe is longer in the model, and is flatter in the image. d4.
Four striations lie on the middle area in the direction parallel to
the Z-axis (long axis) in the shape model but not in the image (Zou
et al. 2014). Moreover, Toutatis' dimensions estimated from the
radar model $(x = 4.60\pm0.10~km, z = 1.92\pm0.10~km)$ is a bit
smaller than those $[(4.75 \times 1.95~km)\pm10\%]$ given by the
spacecraft images (Hudson et al. 2003; Huang et al. 2013a). These
similarities and differences can provide significant improvements to
the current shape model of Toutatis (\cite{Busch14}).

\begin{figure*}
\begin{center}
\includegraphics[scale=0.35]{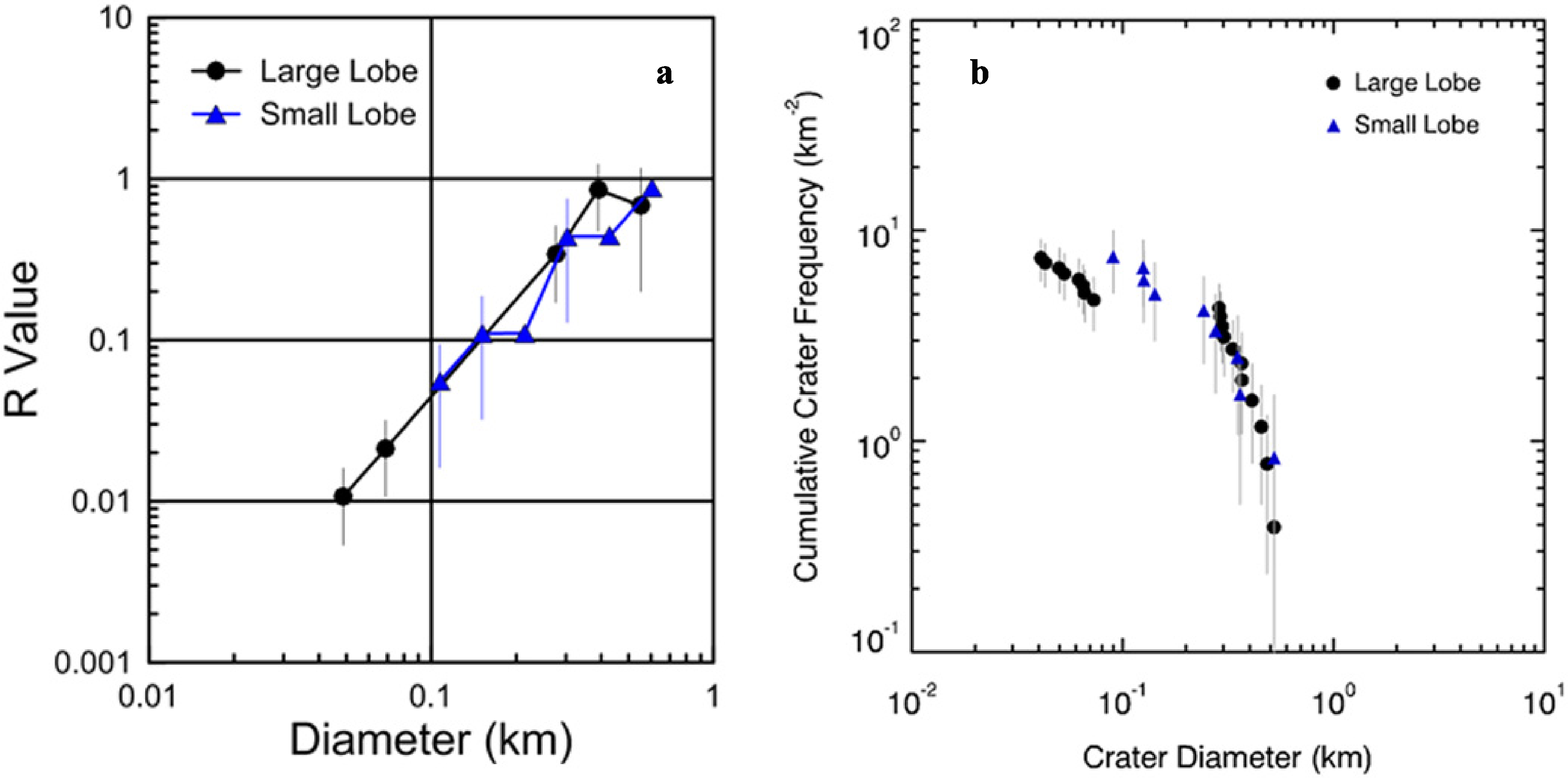}
\caption{The relative (a) and cumulative (b) size-frequency
distribution for craters on Toutatis (Huang et al. 2013a).} \label{geo3}
\end{center}
\end{figure*}

From flyby images, Toutatis is observed to be covered by abundant
concavities, indicating that impact cratering may play an important
role in shaping the present surface. Fresh impact craters have been
defined as ``circular rimmed depressions" (\cite{Melosh2011}).
However, the overall quality of optical images is relatively poor
because they were captured by the monitoring camera. In addition,
the high solar phase angle is not conducive to accurate topographic
analyses. All these factors make the identification of craters on
the imaged side of Toutatis difficult and incompatible for different
researchers. Thus, different numbers of craters have been counted
(Huang et al. 2013a; Zou et al. 2014; Zhu et al. 2014). According to
Huang et al. (2013a), approximately fifty craters have been
identified from 36 to 532 m in size. Most large craters show shallow
depths and obscure shapes, which may result from resetting process
(Chapman et al. 2002). For instance, seismic shaking from subsequent
impacts can cause regolith displacement to erase craters' rims. In
addition, Zhu et al. (2014) regarded the giant depression ($\sim$
800 m) at the big end as a crater. However, Huang et al. (2013a)
suggested that this large-end depression, with relatively subdued
relief, is more likely related to the internal structure of body.
All craters identified from the given images are labeled in blue
profiles in Figure \ref{geo2}. Figure \ref{geo3} shows that the
cumulative size-frequency distributions indicate that two lobes may
undergo either similar history in the cratering (Huang et al.
2013a), or the big lobe may suffer from more impacts than the small
one (Zou et al. 2014). Assuming a solid impactor and proper scaling
law, Zhu et al. (2014) estimated the energy of the impactor for the
800 m depression at the big end to be 5$\times~10^{11}$ J. This
result is fairly greater than the energy required for breaking up a
bulk rock with the same size of Toutatis. Therefore, they inferred
that Toutatis might not bear a monolithic structure but a rubble
pile with fragments accreted. In addition, they calculated the
seismic attenuation factor $\alpha$ = 1.43 for the largest
depression at the big end of Toutatis, which is higher than those of
other porous asteroids. This may greatly attenuate the heavy shock
wave so that abundant large craters are unlikely to lead to global
disruption of Toutatis (Zhu et al. 2014).

\begin{figure*}
\begin{center}
\includegraphics[scale=0.26]{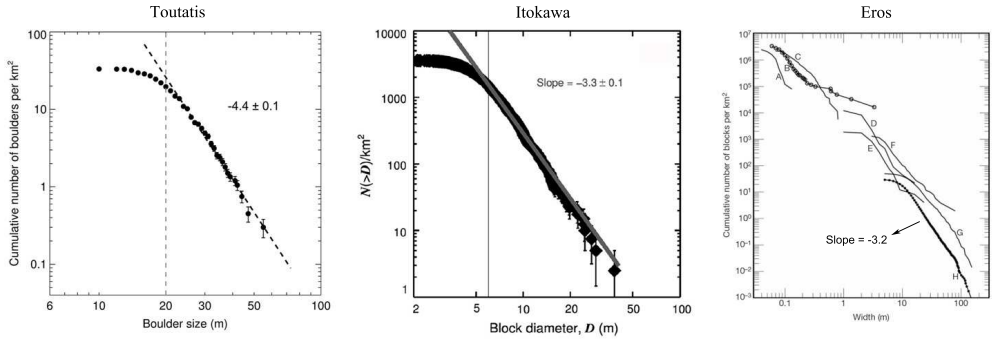}
\caption{A comparative view of cumulative size-frequency
distribution (SFD) of boulders of Toutatis, Itokawa and Eros, which
are reproduced from Jiang et al. (2015a), Mazrouei et al. (2014) and
Thomas et al. (2001), respectively.} \label{boulder}
\end{center}
\end{figure*}

Boulders show a clearly identifiable brightness variation as well as
a bright positive relief with shadow next to it, according to the
criteria of boulders on the surface of Lutetia (K\"{u}ppers et al.
2012). More than 200 boulders scattered across its surface have been
counted over the imaged area of Toutatis (Figure \ref{geo2}). Note
that the boulder size with geometric mean in this work is smaller
than the value with maximum size measured in the previous study
(Huang et al. 2013a; Zou et al. 2014), mainly due to the deviations
in the measurement of boulder size. As a result, boulders have
dimensions ranging from 10 to 61 m, with an average size of 22 m and
90\% of them less than 30 m. The two largest boulders ($>$ 50 m) are
about the neck region. The cumulative boulder size frequency
distribution exhibits a slope of -4.4 $\pm$ 0.1 for 20-60 m size,
which is much steeper than slopes of -3.3 $\pm$ 0.1 for Itokawa
boulders with 6-38 m size and of -3.2 for Eros boulders with 15-80 m
size (Thomas et al. 2001; Mazrouei et al. 2014) (Figure
\ref{boulder}). Moreover, like Itokawa, Toutatis is most likely of
rubble-pile origin, where most boulders are probably fragments from
the parent body but are not generated by impact cratering as for
Eros (Jiang et al. 2015a; see also Jiang et al. 2015b, this volume).
Troughs and ridges are the major types of linear structures observed
from the flyby images. The origin of the troughs on the surface may
arise from the impact of other asteroids.

The optical images of Toutatis acquired by Chang'e-2 spacecraft show
that a handful of small craters have fresh and sharp bowl-shapes,
with upheaval rims without rays, indicative of impacts on loose
regolith. Those large craters with subdued rims and smooth floors
also indicate the existence of fine-grained regolith on the surface
of Toutatis (Huang et al. 2013a; Zhu et al. 2014). \cite[Ostro et
al. (1999)]{Ostro99} surmised that regolith with porosity of lunar
soils should exist and cover nearly one third of the surface of
Toutatis based on their radar observations. As known, the estimation
of the thermal inertia of asteroids' surface can provide a more
effective and reliable way to discern the presence of regolith, as a
low surface thermal inertia generally suggests a regolith surface.
\cite[Howell et al. (1994)]{Howell1994} estimated the thermal
inertia of Toutatis to be $300\sim800\rm~Jm^{-2}K^{-1}s^{-0.5}$ by
employing the thermophysical model to reproduce a $3\rm~\mu m$
spectral region of Toutatis. In addition, using the statistical
relationship between asteroid's thermal inertia and effective
diameter (\cite{Delbo2007}), the surface thermal inertia of Toutatis
can be estimated to be $150\sim225\rm~Jm^{-2}K^{-1}s^{-0.5}$, which
seems to be quite similar to that of Eros
$\sim150\rm~Jm^{-2}K^{-1}s^{-0.5}$, but much lower than that of
Itokawa $\sim750\rm~Jm^{-2}K^{-1}s^{-0.5}$. Although the above
estimation may be rough, it can still work as a reference to infer
the surface situation of the asteroid. As described above, Toutatis
may have a much lower thermal inertia than that of Itokawa, implying
that the coverage of regolith layer over Toutatis' surface is much
more widely than that on Itokawa's surface.

\section{Dynamics and Orientation}

Using the radar data obtained during 1992 flyby,
 \cite[Ostro et al. (1995)]{Ostro95} presented a spin period between
4 and 5 days. \cite[Hudson \& Ostro (1995)]{Hudson95} used a
least-squares estimation to calculate the two major periods, which
were found to be 5.41 days for rotation about the long axis and 7.35
days for precession of the long axis about the axis of angular
momentum vector. Subsequently, Toutatis was observed by Goldstone
during its 1996 approach. \cite[Ostro et al. (1999)]{Ostro99}
further showed that the two periods of Toutatis to be
$5.376\pm0.001$ and $7.420\pm0.005$ days, respectively, by analyzing
the radar measurements. \cite[Scheeres et al. (2000)]{Scheeres2000}
suggested that the tumbling spin state of Toutatis might be a result
of near-Earth flybys over its lifetime. Recently, \cite[Takahashi et
al. (2013)]{Takahashi13} modeled the rotational dynamics and
calculated Toutatis' rotational parameters using radar observations
of five flybys from 1992 to 2008.

Based on the optical results by Chang'e-2, \cite[Zou et al.
(2014)]{Zou14} calculated the imaging distances and image
resolutions, and used profile contour and the similarity method
to measure the attitude angles related to the view port of the
imaging camera, further showed that the contour matching results
are  $\alpha$ = -33.8$^\circ$, $\beta$ = 33.0$^\circ$, $\gamma$= 47.1$^\circ$,
in the form of 3-1-2 Euler angles. According to the graphical frame
described with the angles of direction cos\emph{i} near the flyby epoch,
\cite[Bu et al. (2015)]{Bu15} rotated the radar model and derived the
orientation of 126.13$^\circ\pm$ 0.29$^\circ$, 122.98$^\circ\pm$
0.21$^\circ$ and 126.63$^\circ\pm$ 0.46$^\circ$, without considering
the attitude of the camera in the inertia frame.

Zhao et al. (2015a) employed four frameworks to determine the
orientation of Toutatis at the flyby epoch. Radar-derived models are
rotated about three principal axes to match the optical images. In
combination of the spacecraft's attitude and camera information, the
matching results are adopted to evaluate Toutatis' rotational
parameters through a shooting process. The three of 3-1-3 form Euler
angles from body-fixed frame to inertial coordinate system are
$\alpha=-32^\circ\pm 5^\circ$, $\beta=-50^\circ\pm 0.8^\circ$, and
$\gamma=12^\circ\pm 5^\circ$, respectively.

\begin{figure*}
\centering
\begin{minipage}[c]{0.8\textwidth}
\centering
\includegraphics[scale=0.42]{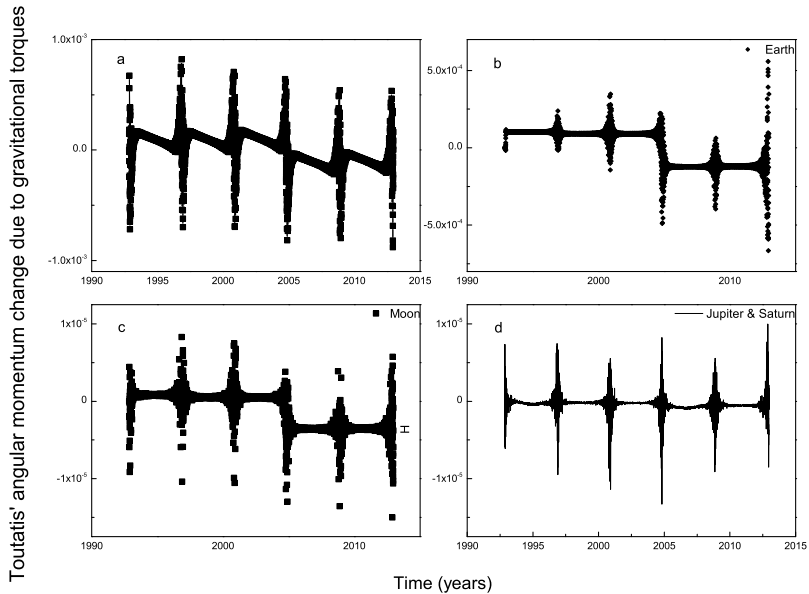}
\end{minipage}%
\caption{Variations in the angular momentum of Toutatis:  (a) all
external gravitational torques, that of  (b) the Earth, (c) the
Moon, (d) Jupiter and Saturn.} \label{moment}
\end{figure*}

Figure \ref{moment} shows that the variations in the angular
momentum originate from all external gravitational torques, and that
of Earth, Moon, and Jupiter and Saturn, respectively
(\cite{Zhao15a}, see also Zhao et al 2015b, this volume). The first
order of external gravitation torque is proven to be ignorable
(\cite{Zhao15a}). Solar tides play a prominent role in all external
perturbations acting on Toutatis' rotational status for the past two
decades. An apparently 1:4 resonance is figured out from the torque
variation curves caused by the Earth and Moon. Regular variations in
Jupiter's tidal effects is generated by the 3:1 mean motion
resonance with Jupiter. Figure \ref{moment} shows the angular
momentum variations that Toutatis experienced due to the Earth and
Moon gravity during the 2004 close approach to Earth. The angular
momentum orientation of Toutatis is determined to be
$\lambda_{H}=180.2^{+0.2^\circ}_{-0.3^\circ}$ and
$\beta_{H}=-54.75^{+0.15^\circ}_{-0.10^\circ}$ . Toutatis is
evaluated to spin along long-axis with a period of 5.38 days and the
long-axis precesses with a period of 7.40 days (\cite{Zhao15a}),
which is in consistence with the previous outcomes
(\cite{Hudson95,Hudson03}).

\section{Formation Scenario}

The Chang'e-2 flyby to Toutatis provides us new observations to
understand the formation scenario of asteroids. As the previous
investigations shown, Toutatis is an S-type asteroid that probably
has a rubble-pile structure. Using the density of L ordinary
chondrites, $\sim$ 3.34 $\rm g~cm^{-3}$ (\cite{Scheeres98}), and
typical density of S-type asteroids, 2.1-2.5 $\rm g~cm^{-3}$
(\cite{Reddy12}), Toutatis' porosity is suggestive of a value
ranging from 25\% to 37\%. This infers that its porosity is between
that of Eros and Itokawa, indicating that Toutatis may be not a
monolith but a coalescence of shattered rocks. Furthermore, such
structures and the bifurcated configuration of Toutatis also imply
that this asteroid is catalogued as a contact binary, which refers
to a sort of asteroid containing two lobes in contact
(\cite{Benner2006}). Several likely origin mechanisms of contact
binaries have been studied (\cite {Cuk07, Sch07, TM11}), and
suggested that two lobes of contact binary may have been separated
once and have bimodal mass distribution (\cite{Benner2006, Bro10}).
The bilobate shape of Toutatis is similar to that of Itokawa and the
nucleus of the 67P/Churyumov-Gerasimenko (\cite{Fujiwara06,
Sierks15}). However, the formation of Toutatis is not fully
understood yet.

Recently, several formation scenarios are proposed to produce this
bilobate shape like Itokawa: (1) two detached objects colliding with
a relatively low speed; (2) two components from the identical parent
body undergoing re-impact and recombination due to YORP and Binary
YORP (Mazrouei et al. 2014);(3) tidal disruption from a terrestrial
planet (Fujiwara et al. 2006) and catastrophic collisions (Huang et
al. 2013a and references therein), etc. As above-mentioned, the
nucleus of 67P/Churyumov-Gerasimenko also consists two lobes, the
body, head and the neck conjunction. There are mainly two scenarios
for the formation of such kind of comet: hierarchical accretion and
accumulation of planetesimals. When it formed,
67P/Churyumov-Gerasimenko is a contact binary which bears a
resemblance to Toutatis (Sierks et al., 2015). Hence, Toutatis, as
well as Itokawa and 67P/Churyumov-Gerasimenko, seems to suffer from
similar formation process.

\section{Future Mission}

Chang'e-2 observations have significantly shed new insights into the
morphological features (such as the large depression in the big end,
the sharply perpendicular profile, the distribution of the craters
and boulders) for  Toutatis. These observations can help us
understand the formation scenarios of this kind of asteroid, which
could provide clues to unveiling the formation of population of
contact binary. Moreover, future Chinese asteroid mission (Multiple
Asteroids Rendezvous and in-situ Survey (MARS) Mission) will
sequently visit three NEOs: (99942)Apophis
(\cite{Binzel09,Souchay14}), and (175706)1996 FG3
(\cite{Pravec98,Pravec00,Mottola00,Yu14,Scheirch15}) are potential
candidates. There are seven scientific payloads proposed on board,
e.g., multi-spectral imager, panoramic camera, penetrating radar,
near infrared spectrometer, gamma-ray spectrometer, in-situ sampling
and analyzer, and ion energy spectrum imager. MARS mission will be
hopeful to provide the key to the formation of planets, the
evolution of Solar system and the origin of life on the Earth.

\acknowledgments{This work is financially supported by National
Natural Science Foundation of China (Grants No. 11273068, 11473073),
the Strategic Priority Research Program-The Emergence of
Cosmological Structures of the Chinese Academy of Sciences (Grant
No. XDB09000000), the innovative and interdisciplinary program by
CAS (Grant No. KJZD-EW-Z001), the Natural Science Foundation of
Jiangsu Province (Grant No. BK20141509), and the Foundation of Minor
Planets of Purple Mountain Observatory. We thank the entire
Chang'e-2 team to make the mission a success.}

\end{document}